\documentclass[prb,aps,twocolumn,floatfix,amsmath,amssymb,tightenlines, showpacs]{revtex4}
\usepackage[pdftex]{graphicx}% Include figure files
 \usepackage{amsmath}
\usepackage{subfigure}
\usepackage[utf8x]{inputenc}
\usepackage[T1]{fontenc}
\usepackage{amssymb}
\usepackage{amsfonts}
\usepackage{bm}
 \usepackage{amsmath} 
 \usepackage[breaklinks=true,colorlinks=true,linkcolor=blue,urlcolor=blue,citecolor=blue]{hyperref}
\usepackage{subfigure}
\usepackage{lipsum}
\usepackage{amsfonts} 
\usepackage{amssymb, mathrsfs}
\usepackage{braket}
\usepackage{graphicx} 
\usepackage[usenames]{color} 

\usepackage{bbm}
\def\beq{\begin{equation}}
\def\eeq{\end{equation}}
\def\bsp{\begin{split}}
\def\esp{\end{split}}
\def\bea{\begin{eqnarray}}
\def\eea{\end{eqnarray}}
\def\ba{\begin{array}}
\def\ea{\end{array}}

\def\lb{\left(}
\def\rb{\right)}

\def\l.{\left.}
\def\r.{\right.}

\def\ra{\rangle}
\def\la{\langle}

\begin{document}

%\date{\today}
\title{Quantum spin Hall effect on the porphyrin lattice}
\author{S. A. Owerre}
\email{sowerre@perimeterinstitute.ca}

\affiliation{Perimeter Institute for Theoretical Physics, 31 Caroline St. N., Waterloo, Ontario N2L 2Y5, Canada.}

%\author{J. Nsofini}
%\email{jnsofini@uwaterloo.ca}
%\affiliation{Institute for Quantum Computing, University of Waterloo, Waterloo, Ontario N2L 3G1, Canada.}
%\affiliation{Department of Physics and Astronomy, University of Waterloo, Waterloo, Ontario N2L 3G1, Canada.}
%

\begin{abstract}
%Topological quantum phase transition in topological insulators is an enthralling phenomena. This phase transition has a unique property in that it is associated with a quantum phase transition point, which separates  different regions with gapped phases, characterized by an immutable topological quantity called the Chern number.

We study the physics of quantum spin Hall (QSH) effect  and topological quantum phase transition on the  porphyrin lattice. We show that in a special limit the pristine model on this lattice  reduces to the usual topological insulator (TI) thin film model. By incorporating explicit time reversal symmetry (TRS) breaking interactions,  we explore the topological properties of this model and show that it exhibits an ordinary insulator phase,  a quantum anomalous Hall (QAH) phase,  and a quantum spin Hall (QSH) phase.
\end{abstract}


\maketitle

\section { Introduction}
The study of topological insulators have become a ubiquitous study  in condensed matter physics. These materials have captivated considerable attention  of researchers \cite{adm, joel, Huichao, burr3, batt, hui, hai1, hui1, yu, hk2, zhang, zhang0, jac, km}. They involve materials such as Bi$_2$Se$_3$ or Bi$_2$Te$_3$. One of the special properties of these materials is that they have an electronic bulk structure  with a finite gap separating the conduction band from the valence band, but their edges (for two-dimensional (2D) TIs) or surfaces (for 3D TIs), however, have gapless states, which are protected by TRS. Provided that the system remains time-reversal invariant, these gapless states are robust. The bulk Hamiltonian near the energy gap closing points (Dirac points) replicates a  2D Dirac Hamiltonian, which describes the surface states.   Interesting electronic transport properties emerge when  TRS is explicitly broken. There are different ways to induce an explicit  TRS breaking interactions. They can be induced by  applying a perpendicular magnetic field or by depositing a ferromagnet   on the surface of a TI.  These TRS breaking interactions lift the degeneracy of the Dirac points and introduce a finite   gap on the surface of a TI. The upshot is that one observes  interesting electronic transport, which include a half quantized conductivity, Majorana bound states, etc.\cite{ sol3, kane, zhang1, adm}. 

However,  the quantum spin Hall (QSH) phase   \cite{hk, batt} is obtained by preserving TRS by the inclusion of an intrinsic spin-orbit coupling (SOC). This phase is reminiscent of two copies of  quantum anomalous Hall (QAH) phases, where each copy breaks TRS and the total system remains time-reversal invariant. It has  been shown that QSH effect can persist even when TRS is explicitly broken \cite{km, yun}. These interesting observations have paved the way for numerous  investigations and the fates of QSH effect in complicated systems.  In the QSH phase, there are two pairs of gapless edge state modes counter-propagating in the vicinity of the bulk gap, which can be captured in the tight binding limit of the low-energy Hamiltonian. The importance of tight binding model in topological insulators was first elucidated by  Bernevig,  Hughes, and Zhang BHZ \cite{batt} on a square lattice. Following the same line of study, Huichao Li {\it et al.} \cite{hui} coined an alternative tight binding model for TI thin film on a square lattice. In recent years, topological insulators have also been investigated on a honeycomb lattice, \cite{Huichao, km, hk, xu}, kagome lattice \cite{guo} Lieb and perovskite lattices \cite{weeks}, and pyrochlore lattice \cite{guo1}.

  The purpose of the paper is to provide another lattice geometry in which the physics of QSH effect and topological quantum phase transition are manifested. We investigate a tight binding model on the porphyrin lattice \cite{joel}, in which  TRS breaking through the Zeeman effect  requires the nearest neighbour sites  to have  a  complex amplitude, reminiscent of the Haldane model \cite{adm}; see Fig.~\eqref{porphyrin_latice}. By incorporating  an additional phase (gauge) into the complex amplitudes we show that in a special limit and with a specific gauge choice, the pristine model reduces to a time-reversal invariant Hamiltonian, which in turn reduces to the usual TI thin film model in the continuum limit.  We also show that this model possesses different Chern numbers in different regimes of the parameter space just like in real systems. We further explore this model by incorporating spin degrees of freedom and explicit TRS breaking couplings, such as an off-diagonal coupling (electrostatic  potential), $t_\parallel$,  a magnetic field , $h$, and a Rashba SOC, $t_R$. We explicitly map out different phase diagrams at the gap closing points.

\section {Tight binding model}
\begin{figure}[ht]
\centering
\includegraphics[width=2.3in]{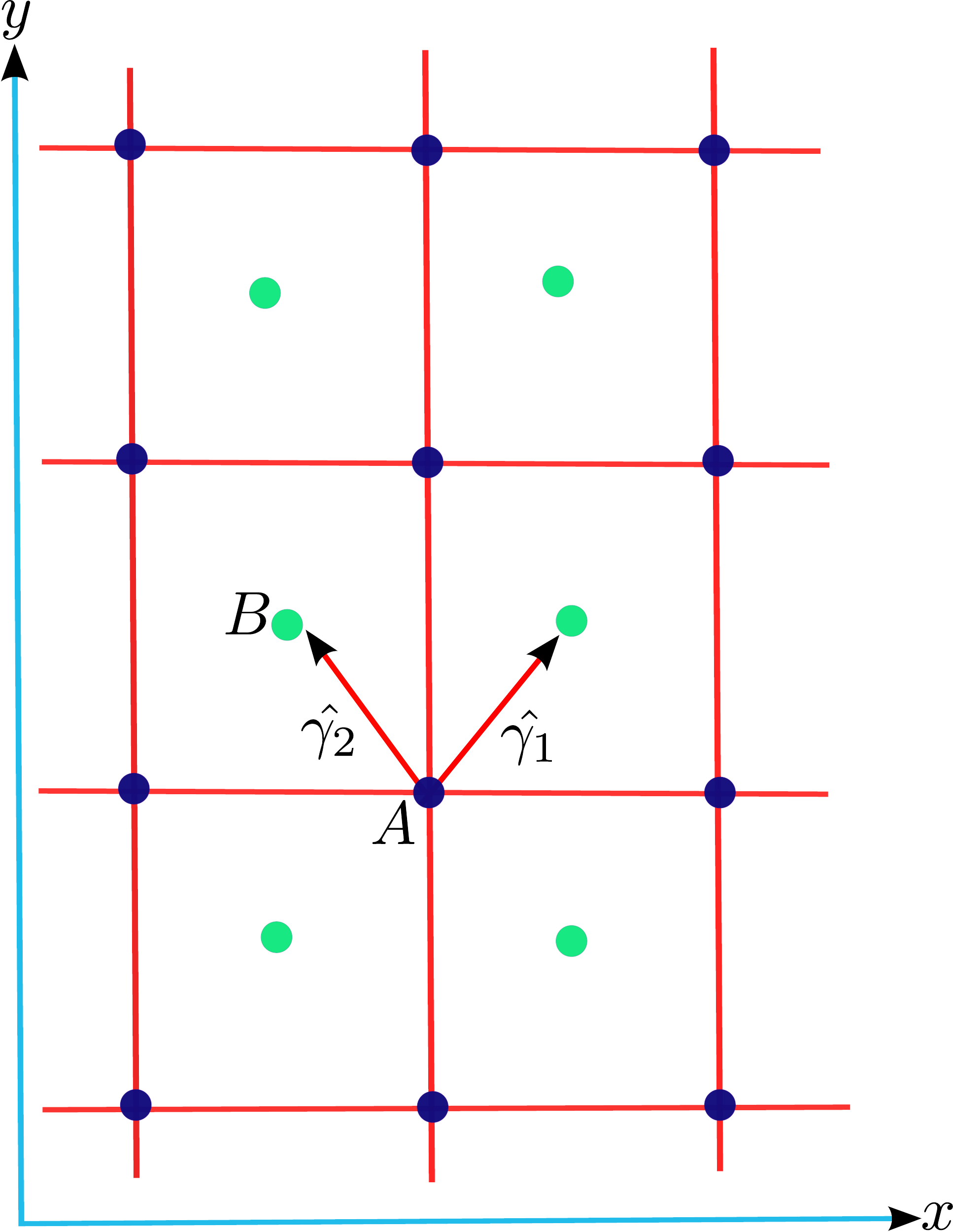}
\caption{Color online.  Porphyrin lattice with two sublattices $A$ and $B$. Nearest neighbour sites are along the diagonals with coordinates $\pm \hat{\gamma}_1$ and $\pm \hat{\gamma}_2$; where $\hat{\gamma_1}= (\hat{x}+\hat y)/2$ and $\hat{\gamma_2}= (-\hat{x}+\hat y)/2$. Next-nearest neighbour sites are along the horizontal and vertical directions  with coordinates $\pm \hat{x}$ and $\pm \hat{y}$ respectively.  }
\label{porphyrin_latice}
\end{figure}

Lattice regularizations of the low-energy Hamiltonian is an expedient  way  for studying the topological properties  and the edge states of a TI in the entire Brillouin zone.  In this paper, we study the tight binding model of porphyrin thin films.  \cite{joel}   The tight binding model can be generally written as \begin{widetext}
\begin{align}
H&=\sum_{\la lm \ra\sigma}\Delta_{lm}e^{i{\Phi_{lm}}} c^\dagger_{l,\sigma}c_{m,\sigma}  +t_m\sum_{l,\sigma\sigma^{\prime}}\tau c^\dagger_{l,\sigma}\lb\tau_z\rb_{\sigma\sigma^{\prime}}c_{l,\sigma^{\prime}}-\frac{t_\perp}{2}\sum_{\la \la lm \ra\ra \sigma\sigma^{\prime}}\tau c^\dagger_{l,\sigma}\lb \tau_z\rb_{\sigma\sigma^{\prime}}c_{m,\sigma^{\prime}} \nonumber\\&-\frac{it_\parallel}{2}\sum_{\la lm \ra\sigma\sigma^{\prime}} \tau c^\dagger_{l,\sigma}c_{m,\sigma^{\prime}}+\frac{it_R}{2}\sum_{\la lm \ra\sigma\sigma^{\prime}}c^\dagger_{l,\sigma}\lb \tau_-\rb_{\sigma\sigma^{\prime}}c_{m,\sigma^{\prime}}+h\sum_{l,\sigma}\tau c^\dagger_{l,\sigma}c_{l,\sigma}.
\label{genbhz}
 \end{align}
 \end{widetext}
Lattice Hamiltonian of this form is usually considered as a toy model used to understand the physics of QSH effect in real materials.\cite{batt, hui, weeks, guo1} 
 Here,  $c^\dagger_{l,\sigma}\thinspace (c_{l,\sigma})$  creates (annihilates) an electron with spin $\sigma \thinspace (\sigma=\uparrow, \downarrow)$ at site $\bold{r}_l$; $\tau=\pm 1$ denotes sublattice $A$ and $B$ respectively; and $\lb \tau_i\rb_{\sigma\sigma^{\prime}}$; where $i=x,y,z$ and $\tau_-=(\tau_x-i\tau_y)/2$ represents the components of sublattice (layer) pseudospins.  $\Phi_{lm}=-\Phi_{ml}=\pm \Phi$ for $m=l\pm \hat{\gamma}_{1,2}$ respectively, {\it i.e.,} if the electron makes a right or left diagonal turn to get to the second bond. The coordinates $ \hat{\gamma}_{1,2}$ are defined in Fig.~\eqref{porphyrin_latice}.  $\Delta_{lm}=\frac{\Delta_{1,2}}{2}$ for $m=l\pm\hat{\gamma}_{1,2}$; where $\Delta_{1,2}$ are complex variables. 
  The summations over $\la lm \ra$ and $\la \la lm \ra\ra$ denote nearest and  next nearest neighbours on a porphyrin square sublattice respectively; see Fig.~\eqref{porphyrin_latice}.
 
In the absence of the  phase $\Phi$ and the spin degrees of freedom, the first three terms in Eq.~\eqref{genbhz} correspond to the model proposed by Joel Yuen-Zhou,	{\it et al.,}\cite{joel} in the context of Frenkel excitons (quasiparticles with no net charge). A special property of this tight binding model is that   the nearest neighbour hopping sites  in Eq.~\eqref{genbhz}  have complex amplitudes, which are explained in Ref.~[\onlinecite{joel}]. The motivation for the phase factor (gauge) $\Phi$ is rather technical. It does not, however, change the complex nature of the hopping amplitudes. In particular,  the total phase in a unit cell vanishes, reminiscent of the Haldane model. \cite{adm}  The reason for this phase factor will become apparent in the subsequent sections, where we show that different choices of $\Phi$ lead to different surface Hamiltonians, perhaps not different physics.  The second and the third terms  represent the diagonal couplings between the upper and the lower surface states of the thin film.  In addition, we have introduced other couplings in order to enhance the topological properties of this thin film.    The fourth term is an electrostatic  potential which couples (off-diagonal) the upper and lower surface states;  the fifth term can be identified as the Rashba-like SOC in the thin film, and the last term is a staggered sublattice potential, which corresponds to a magnetic field in the thin film.

Next, it is expedient to Fourier transform into momentum space, {\it i.e.}, $c_l=\frac{1}{\sqrt{N}}\sum_{\bold k} c_\bold{k} e^{-\bold{k}\cdot l}$.  We obtain 
 \bea
 H=\sum_{\bold k} \psi^\dagger(\bold{k})\mathcal{H}(\bold{k})\psi(\bold{k}),
 \eea  
 where  $\psi^\dagger(\bold{k})=\lb c^\dagger_{\bold{k} A,\uparrow} \thinspace c^\dagger_{\bold{k} B,\uparrow} \thinspace c^\dagger_{\bold{k} A, \downarrow}\thinspace c^\dagger_{\bold{k} B, \downarrow}\rb$ is the Nambu operators  for the thin film.
 %\begin{align}
% H&= \sum_{\bold{Q},\sigma}c^\dagger_{\bold{Q}, \sigma}\bigg[ v_F\lb \sigma_x\sin k_y - \sigma_y\sin k_x\rb -\sigmat_{\perp}\sigma_z\bigg] c_{\bold{Q}, \sigma}\nonumber\\&-t_{\parallel}\sum_{\bold{Q}}\lb \cos k_x +\cos k_y \rb\lb c^\dagger_{\bold{Q},\uparrow} \sigma_x c_{\bold{Q},\downarrow} +h.c\rb,
% \end{align}
%which can be written as

\begin{widetext}
\begin{align}
\mathcal{H}(\bold k)&= \bigg[\rho_1\cos\lb\frac{k_x+k_y}{2}-{\Phi}\rb+\rho_2\cos\lb\frac{k_x-k_y}{2}+{\Phi}\rb\bigg]\bold{I}_\tau\sigma_x -\bigg[\bar{\rho_1}\cos\lb\frac{k_x+k_y}{2}-{\Phi}\rb+\bar{\rho_2}\cos\lb\frac{k_x-k_y}{2}+{\Phi}\rb\bigg]\bold{I}_\tau\sigma_y \nonumber\\&+[{t_m}-t_{\perp}\lb  \cos k_x+\cos k_y\rb ]\tau_z \sigma_z+ 2t_{\parallel}\lb \cos k_x \cos k_y \rb\tau_x \sigma_y +2t_{R}\cos k_x \cos k_y \lb \tau_y \sigma_x -\tau_x \sigma_y\rb +h\bold{I}_\tau\sigma_z; 
\label{fullti}
\end{align}
\end{widetext}
where $\rho_1=\mathfrak{R}\Delta_1$, $\rho_2=\mathfrak{R}\Delta_2$;  and $\bar{\rho_1}=\mathfrak{I}\Delta_1$, $\bar{\rho_2}=\mathfrak{I}\Delta_2$; $\mathfrak{R}$  and $\mathfrak{I}$ denote real and imaginary parts. 
The Pauli matrices $\sigma_i$  act on the real spin space, while $\tau_i$ act on the upper and the lower surfaces of the thin film,  with spin-up on the upper surface and spin-down on the lower one and $\bold{I}_\tau$ is a $2\times 2$ identity matrix.  

\section { Quantum spin Hall state}
In order the see how the physics of QSH effect is manifested on the porphyrin lattice, we consider a special limit of Eq.~\eqref{fullti}. This limit corresponds to the first three terms in Eq.~\eqref{fullti} ({\it i.e.,} all the terms involving the big and small square brackets), and we call it the {\it pristine  model}.   The  first two terms (big square brackets) in Eq.~\eqref{fullti} yield different functions depending on the choice of $\Phi$.  The gauge choice $\Phi=0$ amounts to the Hamiltonian  studied in Ref.~[\onlinecite{joel}].  We consider an alternative gauge choice of the pristine Hamiltonian with $\Phi=\pi/2$. In this gauge, the Hamiltonian decouples into two TRS copies, {\it i.e.,}$\mathcal{T}$: $\mathcal{H}_\uparrow \mapsto \mathcal{H}_\downarrow$, where the time reversal operator is given by $\mathcal{T}= i\sigma_y\mathcal{K}$ and $\mathcal{K}$ is a complex conjugation. Next, we take the limit $\rho_1=\rho_2=\rho$ and  $\bar{\rho_1}=-\bar{\rho_2}=\rho$, which corresponds to $\Delta_1=\Delta_2^*$. The upshot of these limits is that the Hamiltonian simplifies as:

\begin{align}
\mathcal{H}(\bold k)&= 2\rho\cos\lb\frac{k_x}{2}\rb\sin\lb\frac{k_y}{2}\rb \bold{I}_\tau\sigma_x\nonumber\\&\label{thinfilm1}
 - 2{\rho}\cos\lb\frac{k_y}{2}\rb\sin\lb\frac{k_x}{2}\rb \bold{I}_\tau\sigma_y\\& \nonumber +[t_m-t_{\perp}\lb \cos k_x+\cos k_y\rb]\tau_z \sigma_z.
\end{align}
Expanding Eq.~\eqref{thinfilm1} near the $\Gamma$ point $(\bold k=0)$, we see that the resulting low-energy Hamiltonian is analogous to that of a 3D TI thin film. \cite{hui, hai1, hui1, Huichao, burr3} 
\begin{figure}[ht]
\centering
\includegraphics[width=2.3in]{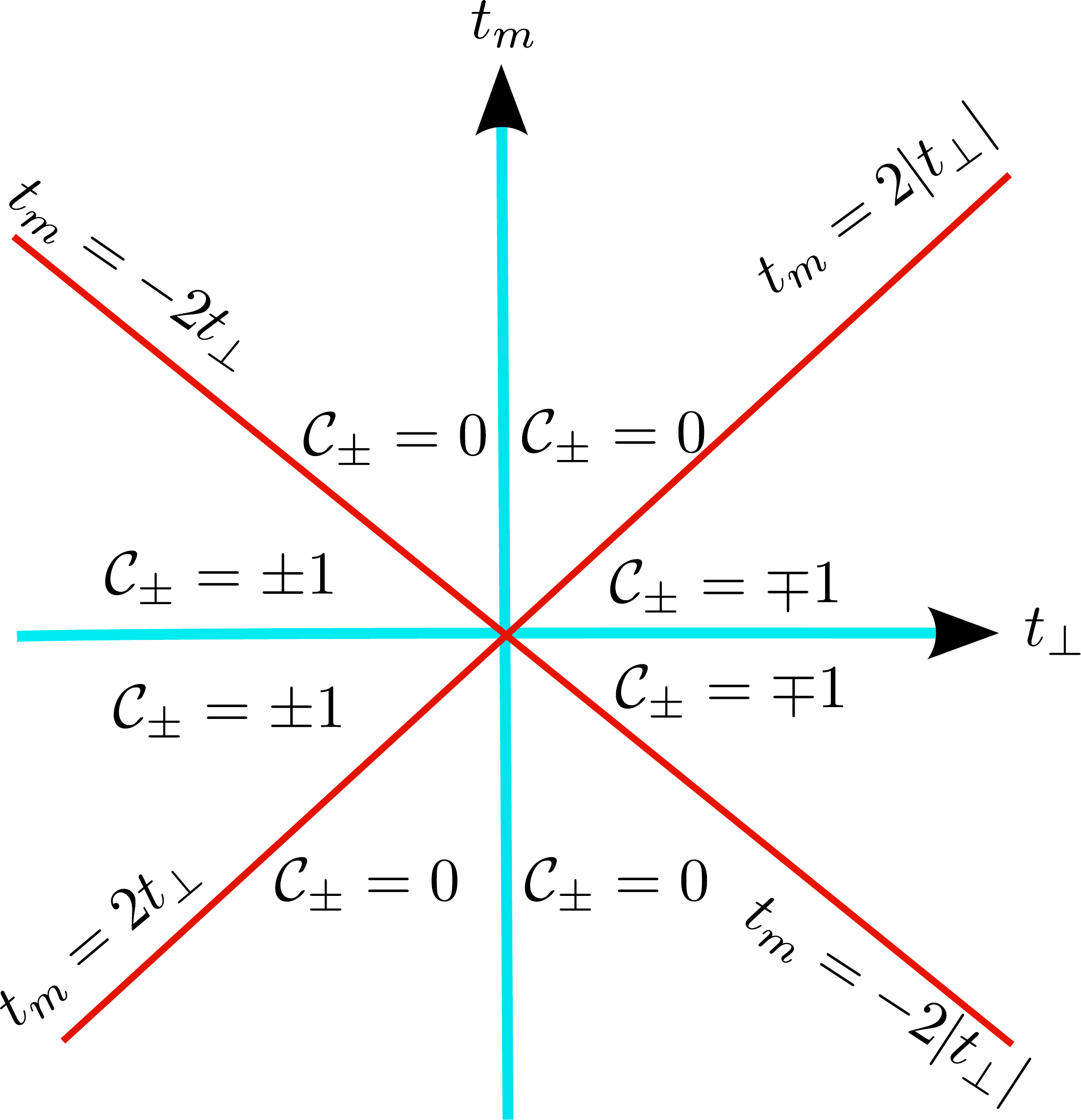}
\caption{Color online.  The phase diagram of the pristine model. The phase boundary $ t_m =\pm 2|t_\perp|$ separates different phases with trivial and non-trivial Chern numbers. }
\label{TI_phaseb3}
\end{figure}
\begin{figure}[ht]
\centering
\includegraphics[width=3in]{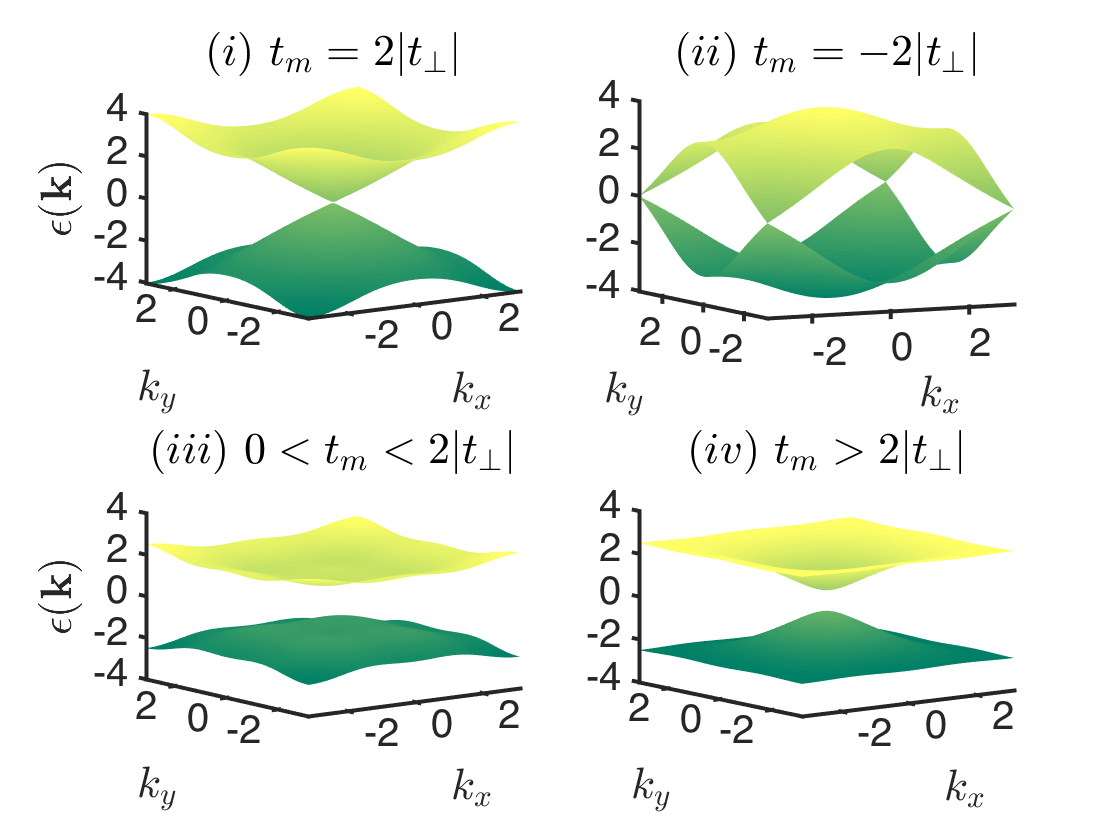}
\caption{Color online. The spin up energy bands of the tight binding model in Eq.~\eqref{thinfilm1} showing the metallic phases $(i)$ and $(ii)$,  the QSH phase $(iii)$, and the insulating phase $(iv)$. }
\label{Pris}
\end{figure}
As the pristine Hamiltonian, Eq.~\eqref{thinfilm1}, has a different structure from previously studied model on the square lattice,\cite{batt,hui} we will show how topological effects are manifested in Eq.~\eqref{thinfilm1} just like in the previously studied models.  As the system possesses TRS, the corresponding eigenvalues are two-fold degenerate.  The topological phase transition points are located at the momenta where the energy eigenvalues vanish. There are four semi-metallic states at the time reversal invariant momenta (TRIM) points:   $\bold{Q}_0=(0,0)$ and  $\bold{Q}_1=(\pi, \pi)$ for $t_m=\pm 2t_\perp$ respectively (see Fig.~\eqref{Pris}) and $\bold{Q}_2=(0,\pi)$ and  $\bold{Q}_3=(\pi, 0)$ for $t_m=0$ respectively . We focus on the points $\bold{Q}_{0}$ and $\bold{Q}_{1}$, with  nontrivial phase boundaries. The phase boundaries separate regions with QSH phase and ordinary insulator phase.   The spin Chern number characterizing these different phases is given by \cite{thou}
\begin{align}
\mathcal{C}=\frac{1}{2\pi}\sum_{l}\int d^2k\thinspace \Omega_l.
\label{chern1}
\end{align}

The Berry curvature $\Omega_l(k)$  has the form:
\begin{align}
\Omega_l(\bold k)=-\sum_{l\neq m}\frac{2\text{Im}\lb \braket{\psi_{l\bold k}|v_x|\psi_{m\bold k}}\braket{\psi_{m\bold k}|v_y|\psi_{l\bold k}}\rb}{\lb\epsilon_l-\epsilon_m\rb^2},
\label{chern2}
\end{align}
where the summation is over all occupied valence bands in the first Brillouin zone below the bulk energy gap, and  $v_{i}=\partial \mathcal{H}(\bold k)/\partial k_{i}$; $i=x,y$ define the velocity operators.
 \begin{figure}[ht]
\centering
\includegraphics[width=3in]{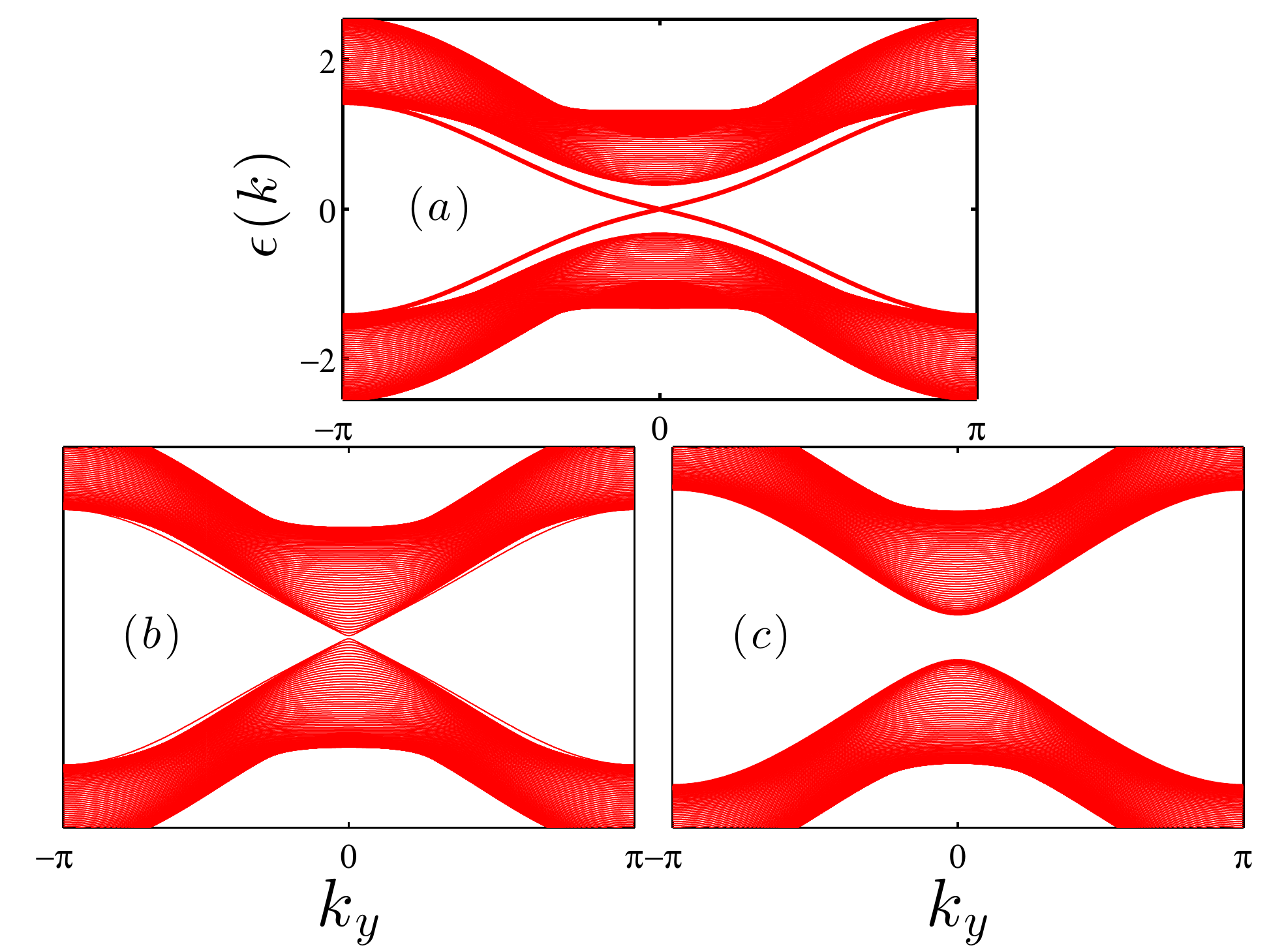}
\caption{Color online.  The bulk energy spectrum and the edge states for the pristine model along the $k_y$ direction, with $t_\perp=1$ and $\rho=1$.  $(a)$ $t_m<2|t_\perp|$;  $(b)$ $t_m=2|t_\perp|$;  $(c)$ $ t_m >2|t_\perp|$. }
\label{QSH}
\end{figure}
The Chern numbers can be computed separately. The appropriate formula is given by \cite{vol}

\bea
 \mathcal{C}_{\tau_z}=\frac{1}{4\pi}\int dk_xdk_y\frac{\epsilon_{\alpha\beta\gamma}}{n^3}\frac{\partial n_\alpha}{\partial k_x}\frac{\partial n_\beta}{\partial k_y}n_{\gamma},
\label{hall}
\eea
 
where $\epsilon_{\alpha\beta\gamma}$ is an antisymmetric tensor; $n_i$ are the components of a unit vector with, 
\begin{align}
 &n_x=2\rho\cos\lb\frac{k_x}{2}\rb\sin\lb\frac{k_y}{2}\rb; \\& n_y=-2{\rho}\cos\lb\frac{k_y}{2}\rb\sin\lb\frac{k_x}{2}\rb; \\&n_z=( t_m-t_{\perp}\lb\cos k_x+\cos k_y\rb)\tau_z.
 \end{align}
  For this model, all the six terms in Eq.~\eqref{hall} contribute equally. The numerical computation of the spin Chern numbers in the entire Brillouin zone yields
\begin{align}
 \mathcal{C}_{\pm}&= 
  \begin{cases}
  0, \quad  t_m \in \lb-\infty, -2|t_\perp|\rb \thinspace \text{and} \thinspace  t_m \in \lb 2|t_\perp|,\infty \rb;\\  \thinspace \text{undefined} \thinspace, \quad   t_m =\pm 2|t_\perp|;\\\mp 1, \quad t_m \in \lb -2|t_\perp|, 2|t_\perp| \rb;   \end{cases}
  \label{chern}
  \end{align}
independent of $\rho$.  The  trivial Chern number corresponds to an  ordinary insulator phase.  The Chern number is ill-defined at the  gap closing points (semi-metallic phase). The gap further reopens, which corresponds to a QSH phase  with a nontrivial Chern number. The phase diagram is depicted in Fig.~\eqref{TI_phaseb3}, where the phase boundary $ t_m =\pm 2|t_\perp|$ separates different phases with trivial and non-trivial Chern numbers.

The nontrivial Chern numbers in the QSH phase indicate that there are a pair of counter-propagating  edge states in the vicinity of the bulk gap. We have solved for the edge states by diagonalizing the Hamiltonian in a cylindrical geometry, with periodic boundary conditions along the $y$ direction and open conditions along the $x$ direction  \cite{zhang1}.   In Fig.~\eqref{QSH}, we show the plot of the bulk energy spectrum in the three phases shown in Eq.~\eqref{chern}.  The edge states clearly reveal the topological properties of this system.  It is apparent that in the QSH regime $(a)$, there are two pairs  of edge states modes with opposite pseudospins (which are degenerate)  that propagate in the vicinity of the bulk gap; these states are gapless, $\epsilon(k_y)=0$ at  $k_y=0$, corresponding to the absolute value of the Chern number \cite{hat}.  The bulk gap closes at $(b)$ corresponding to the phase transition point. In the insulating regime, the gap reopens and  the edge states participate in the bulk gap separation  $(c)$, with no counter-propagating edge states modes in the vicinity of the bulk gap.

\section { TRS breaking interactions}

In this section we study the topological phases of the full Hamiltonian in Eq.~\eqref{fullti}. Henceforth we work with the gauge \cite{gauge} $\Phi=\pi/2$ and take $\Delta_1=\Delta_2^*$, which corresponds to $\rho_1=\rho_2=\rho$ and  $\bar{\rho_1}=-\bar{\rho_2}=\rho$. We further take $\rho$ as the energy unit. The full Hamiltonian in Eq.~\eqref{fullti} does not possess any analytical diagonalization.   We find that at TRIM points $\bold{Q}_{0}$ and $\bold{Q}_{1}$ the topological phase boundary is given by
\begin{align}
h_c=\pm |{t}^{s}|\sqrt{\bigg[ 1-\lb\frac{2t_R}{ {t}^{s}}\rb^2\bigg] \bigg[ 1-\lb\frac{2 \bar{t}_R}{ {t}^{s}}\rb^2\bigg] },
\label{crit}
\end{align}
\begin{figure}[ht]
  \centering
 \includegraphics[width=3.2in]{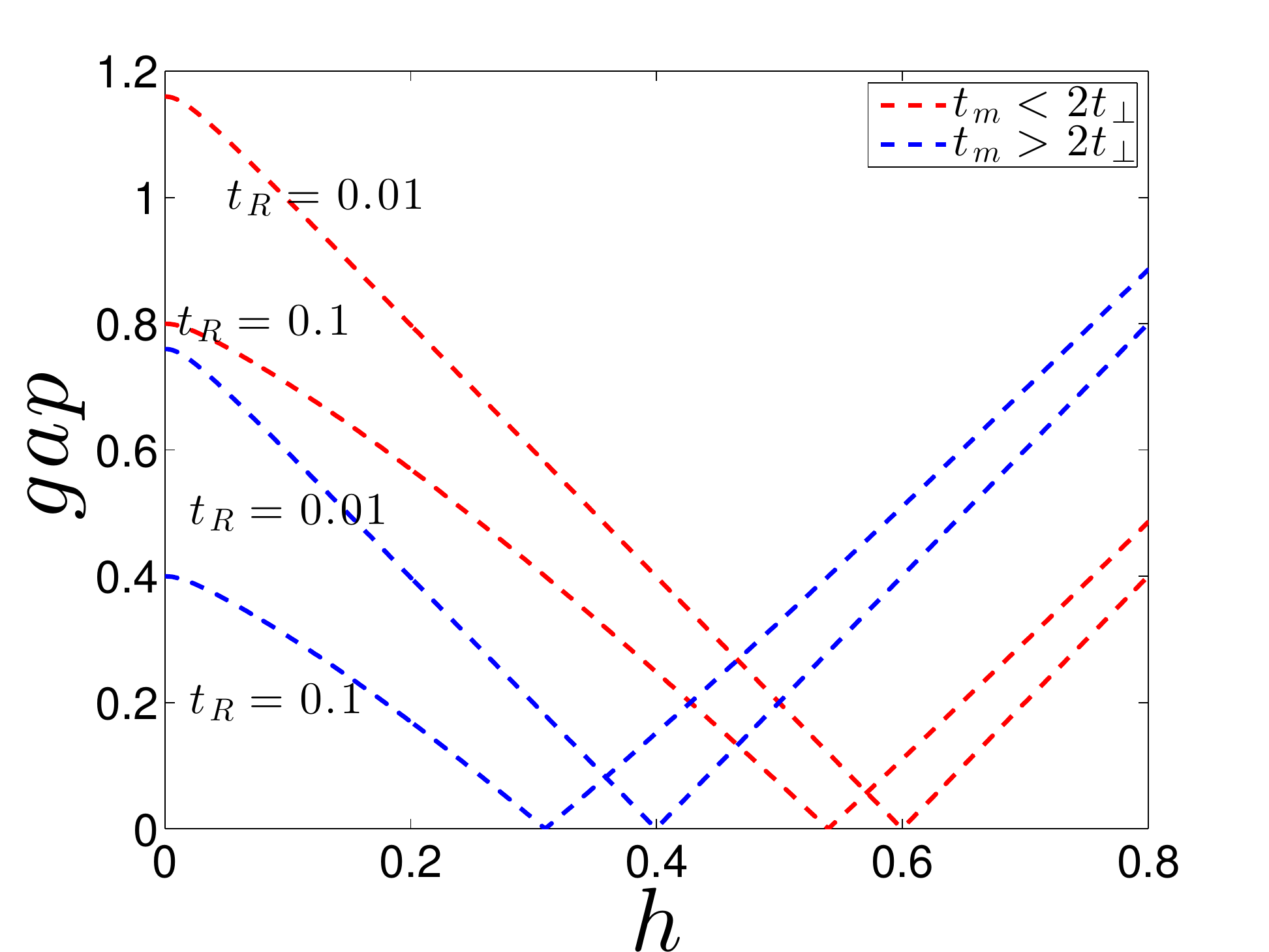}
 \caption{Color online.  The bulk energy gap for the low energy bands of the full Hamiltonian at $\bold{Q}_0$ as a function of $h$ for fixed value of $t_\perp=0.8$, $t_\parallel=0.01$. The red dash line is for  $t_m=1$ and the blue dash line is for  $t_m=2$. There are two gapped regions separated by a gap closing point in the QSH regime $t_m<2|t_\perp|$ and the ordinary insulating regime $t_m> 2|t_\perp|$ of the pristine model. }
 \label{energy_gap}
 \end{figure}
\begin{figure*}[ht]
\centering
\includegraphics[width=5in]{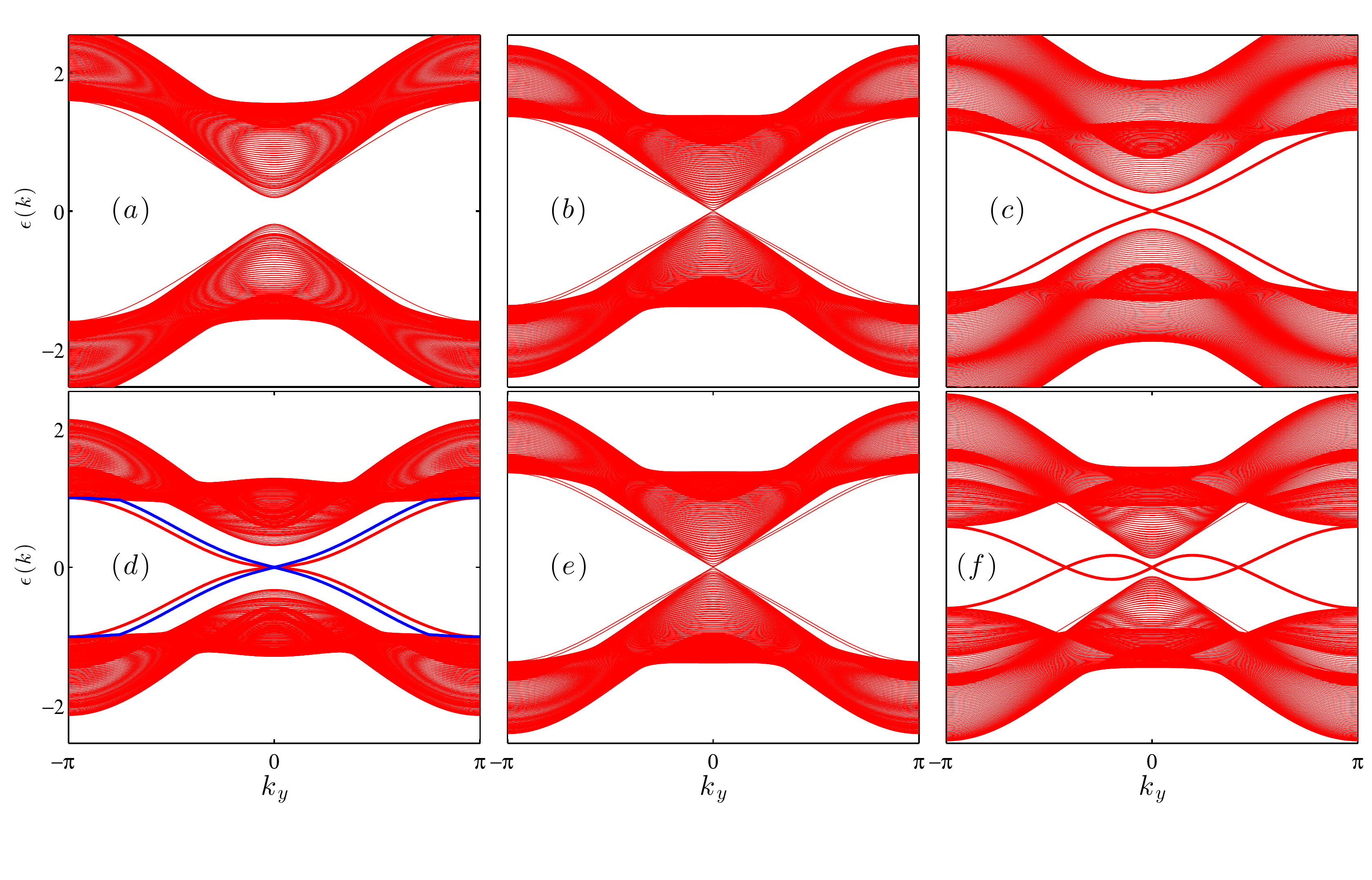}
\caption{Color online.  The bulk energy spectrum and the edge states for the full tight binding model along the $k_y$ direction. The parameters are:  $(a)\thinspace h=0.1$ $(b)\thinspace h= h_c$ $(c)\thinspace h=0.8$ with $ t_m>2|t_\perp|$ and  $(d)\thinspace h=0.1$ $(e)\thinspace h= h_c$ $(f)\thinspace h=0.8$ with $t_m<2|t_\perp|$. Other parameters are the same as in Fig.~\eqref{energy_gap} with  $t_R=0.01$. See text for explanation.}
\label{Full_TI_edge}
\end{figure*}
which is determined by the condition that the conduction and valence bands touch each other as in the case of graphene \cite{km, yun}. Here,  $\bar{t}_R={t_R} +t_\parallel$;  $t^{s}= t_m +2st_\perp$, and $s=\mp$ at $\bold{Q}_{0}$ and $\bold{Q}_{1}$ respectively. The phase boundary at the other TRIM points $\bold{Q}_{2}$ and $\bold{Q}_{3}$  corresponds to $t_\perp=0$ in Eq.~\eqref{crit}.   Note that the topological phase boundary diverges at the phase boundary of the pristine model $t_m=\pm 2t_\perp$. Now we consider  the QSH regime $t_m< 2|t_\perp|$ and the ordinary insulating regime $t_m> 2|t_\perp|$ of the pristine model, with  fixed value of $t_\parallel$, and several values of $t_R$.   We find that the bulk gap as a function of $h$ at $\bold{Q}_{0}$ initially increases at $h=0$, and gradually decreases as $h$ increases from zero, and then vanishes at the topological phase boundary $h_c$.  The  gap further reopens for $h>h_c$; see Fig.~\eqref{energy_gap}. This is not different from what is observed in other lattice geometries such as the honeycomb lattice, \cite{Huichao, km, hk, xu}, the kagome lattice \cite{guo}, the Lieb and perovskite lattices \cite{weeks}, and the pyrochlore lattice \cite{guo1}.

\begin{figure}[ht]
\centering
\includegraphics[width=2.3in]{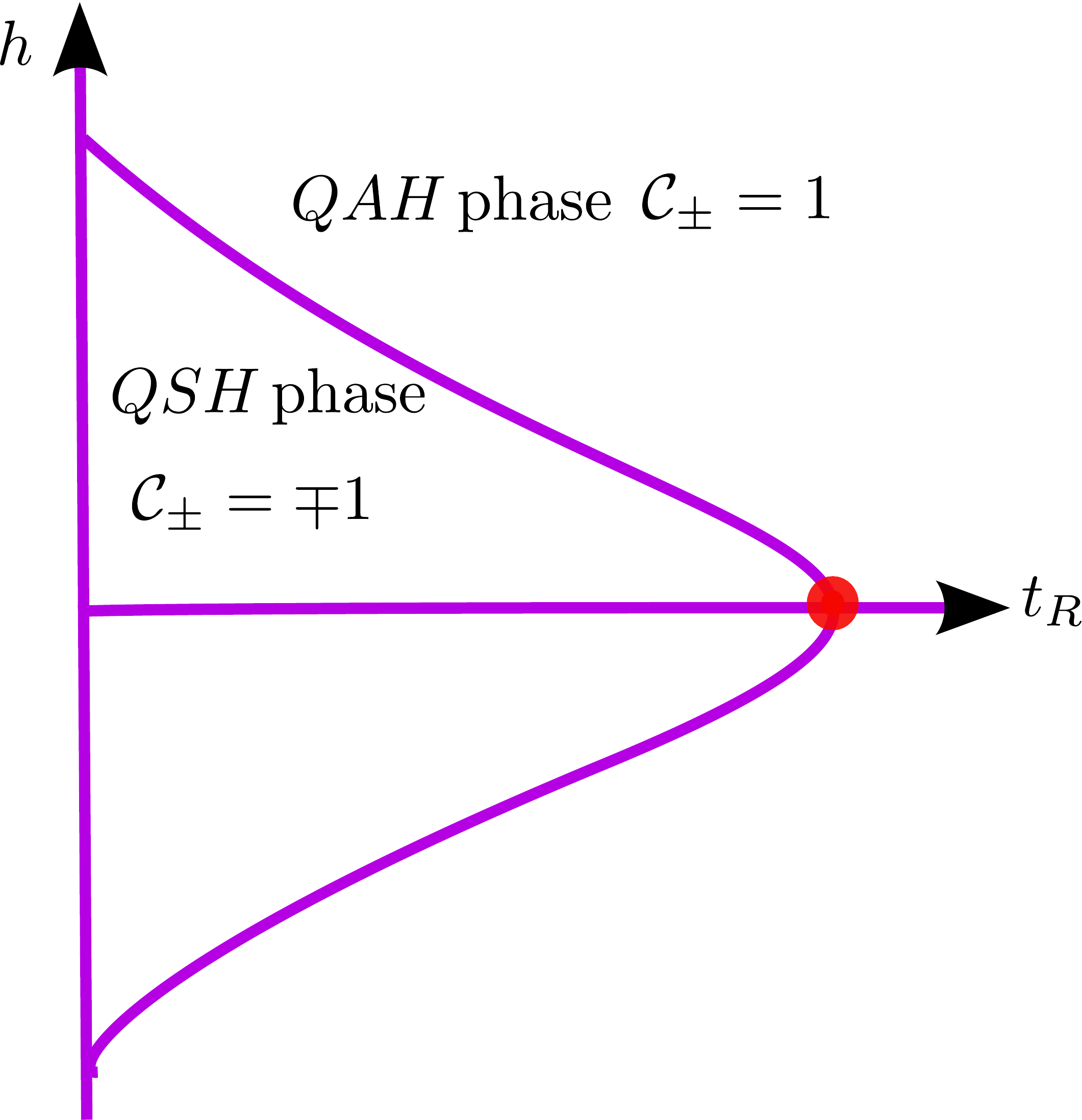}
\caption{Color online.  The phase diagram shown for fixed  values of $t_\parallel>0$ and $t_{\perp}>0$ with $t_{m}<2|t_{\perp}|$. The red dot represents the point  where the boundary vanishes. }
\label{TI_phaseb2}
\end{figure}

 We have solved for the edge states to substantiate the topological properties of these phases. In Fig.~\eqref{Full_TI_edge}, we plot the edge states along the $k_{y}$ direction.  In the ordinary insulating regime ($t_{m}>2|t_{\perp}|$) of the pristine model with $t_\perp=0.8; ~  t_\parallel=t_R=0.01$  $(a)-(c)$,  we see that there is an ordinary insulating phase with no edge state modes at $h<h_c$. The gap closes at the phase boundary $h=h_{c}$ which indicates a phase transition. 
As seen in Fig.~\eqref{Full_TI_edge}, at $h>h_{c}$ the magnetic field dominates,  we see that the gap reopens and there are a pair of counter-propagating edge state modes in the vicinity of the bulk gap, which indicates a QAH phase. We find that  no QSH phase appears in this regime. On the other hand, in the QSH regime ($t_{m}<2|t_{\perp}|$) of the pristine model with $t_\perp=0.8;~   t_\parallel=t_R=0.01$  $(d)-(f)$, there are two pairs of counter-propagating edge state modes in the vicinity of the bulk gap at $h<h_c$, where the red and blue colours represent the edge state modes with opposite pseudospins.  Again the gap closes at the phase boundary $h=h_{c}$ which indicates a phase transition. At $h>h_{c}$, the magnetic field once again dominates and the gap reopens with  a pair of counter-propagating edge state modes in the vicinity of the bulk gap, indicating a QAH phase. On the contrary, no ordinary insulator phase appears in this regime.  The phase diagram in this regime is shown in Fig.~\eqref{TI_phaseb2}.    It is interesting to note that a QAH phase appears in each regime of the pristine model. We have checked that  the continuum Hamiltonian also corroborates our results. Thus, the  porphyrin lattice provides another lattice geometry to study QSH effect and topological quantum phase transition. 

\section {Conclusion}
The  porphyrin lattice represents another lattice geometry where the physics of quantum spin Hall (QSH) effect and topological quantum phase transition are manifested. In this paper, we have explicitly shown how these physics can be constructed from a tight binding toy model.  We showed that the porphyrin lattice tight binding model captures the topological quantum phase transitions observed in other lattice geometries. In particular, the transitions between  QSH phase and ordinary insulator phase, as well as quantum anomalous Hall (QAH) phase.  We also showed that the complex hopping amplitudes and the  gauge degree of freedom are the distinguishing features of the porphyrin lattice tight binding model as was first introduced in  Ref.~[\onlinecite{joel}].

\section {Acknowledgments}
The author is indebted to Tami Pereg-Barnea for pointing out a crucial error in a related model studied by the author. Research at Perimeter Institute is supported by the Government of Canada through Industry Canada and by the Province of Ontario through the Ministry of Research
and Innovation.


\begin{thebibliography}{99}
\bibitem{gauge}
There are several gauge choices which give different surface Hamiltonians. Time-reversal invariant pristine Hamiltonian is obtained when $\Phi=(2n+1)\pi/2$, where $n\in\mathbb{Z}$.
\end{thebibliography}
\end{document}